\documentclass{elsart5p}

\usepackage{graphics}
\usepackage{graphicx}
\usepackage{amssymb}
\begin{document}

\begin{frontmatter}



\title{Direct observation of quantum superconducting fluctuations in an insulating groundstate}
%

\author[AA]{N.P. Armitage\corauthref{NPA}},
\ead{npa@pha.jhu.edu}
\author[BB]{R. W. Crane},
\author[CC]{G. Sambandamurthy}
\author[DD]{A. Johansson}
\author[DD]{D. Shahar}
\author[EE]{V. Zaretskey}  
\author[EE]{G. Gr\"{u}ner}

\address[AA]{Department of Physics and Astronomy, The Johns Hopkins University, Baltimore, MD 21218, USA }
\address[BB]{Chair of Entrepreneurial Risks, Kreuzplatz 5, KPL H22, CH-8032 Zurich, Switzerland}
\address[CC]{SUNY - Buffalo, Department of Physics, 239 Fronczak Hall, Buffalo, NY 14260-1500, USA}
\address[DD]{Department of Condensed Matter Physics, Weizmann Institute of Science, Rehovot 76100, Israel}
\address[EE]{Department of Physics and Astronomy, University of California, Los Angeles, CA 90095, USA}

\corauth[NPA]{Corresponding author }

\begin{abstract}

We review our recent measurements of the complex AC conductivity of thin InO$_x$ films studied as a function of magnetic field through the nominal 2D superconductor-insulator transition.  These measurements - the first of their type to probe nonzero frequency - reveals a significant finite frequency superfluid stiffness well into the insulating regime.  Unlike conventional fluctuation superconductivity in which $thermal$ fluctuations give a superconducting response in regions of parameter space that don't exhibit long range order, these fluctuations are temperature independent as $T \rightarrow 0$ and are exhibited in samples where the resistance is large  (greater than $10^6 \Omega/ \Box$) and strongly diverging.  We interpret this as the direct observation of quantum superconducting fluctuations around an insulating ground state.  This system serves as a prototype for other insulating states of matter that derive from superconductors.
\end{abstract}

\begin{keyword}
Quantum fluctuations; Quantum phase transition;  2D Superconductor-Insulator; AC conductivity
\PACS 78.67.-n,74.78.-w, 74.25.Gz, 74.25.Dw, 73.43.Nq
\end{keyword}

\end{frontmatter}

The T=0 2D superconductor-insulator transition  \cite{Nina,Paalanenscaling} is a particularly beautiful and illustrative quantum phase transition which reflects a transition between the two eigenstates at the extremes of a superconductor's uncertainty relation between phase and particle number ($ \Delta \theta \Delta n> 1$).  One of the main impediments to its detailed understanding has been the dearth of applied experimental probes other than DC transport and tunneling.  In this regard, we have succeeded recently in performing a comprehensive AC conductivity study across the superconductor-insulator transition \cite{Crane0Field,CraneFiniteField}.  These microwave measurements are in the important $\hbar \omega >> k_B T$ limit \cite{Sondhi} and have resolved a substantial superfluid stiffness well into the insulating state.

Samples were 200 $\AA$-thick 3mm-diameter highly-disordered amorphous indium oxide ($\alpha$:InO) thin films prepared by e-gun evaporating high purity (99.999\%) In$_2$O$_3$ on clean sapphire discs in high vacuum.   All experimental measures show them to be homogeneous down to the nanometer scale.  Details of their growth and characterization can be found elsewhere \cite{Crane0Field,CraneFiniteField,Kowal,Murthy04,Steiner2}. Complex conductivity measurements were performed in a novel cryomagnetic resonant microwave cavity system at a number of discrete frequencies ranging from 9 to 110 GHz.  Details of the measurement apparatus and our analysis scheme can also be found elsewhere \cite{Crane0Field,CraneFiniteField}.

At low temperatures, the imaginary conductivity for a long-range ordered superconductor is expected to have the form $\sigma_{2} = \frac{N e^2}{\omega m}$ where $N$ is the superfluid density and $e$ and $m$ are the electronic charge and mass.  For a fluctuating superconductor one can define $ \sigma_2 = \frac{N(\omega) e^2}{\omega m}$  where an additional frequency dependence is captured by a generalized frequency dependent superfluid density. The superfluid density is directly proportional to the superfluid stiffness which is the energy scale for inducing slips in the superconducting phase.  In Fig. 1 we display the field and temperature dependence of the generalized finite frequency superfluid stiffness measured at 22 GHz, $T_{\theta}$ (in degrees Kelvin) extracted via the relation $\sigma_2 = \sigma_Q \frac{k_B T_{\theta}}{\hbar \omega}$, where $\sigma_Q = \frac{4e^2}{hd} $ is the quantum of conductance for Cooper pairs divided by the film thickness.   Similar plots can be made at other frequencies.    A discussion of our neglect of the normal electron contribution to $\sigma_2$ can be found elsewhere \cite{Crane0Field,CraneFiniteField}.  The finite-frequency superfluid stiffness falls quickly with increasing field, but remains finite above $H_{SIT}$ and well into the insulating regime to fields almost 3 times the critical field H$_{SIT} = 3.68 T$.  Here the critical field  has been defined by the iso-resistance point from DC measurements that were performed concurrently \cite{CraneFiniteField}.  This is the first direct model-free measure of superconducting correlations on the insulating side of the 2D superconductor-insulator transition in an amorphous film.  It is the use of relatively high probing frequencies that allows us to resolve superconducting fluctuations into the insulating part of the phase diagram.

\begin{figure}[!ht]
\begin{center}
\includegraphics[angle=-0,width=0.4\textwidth]{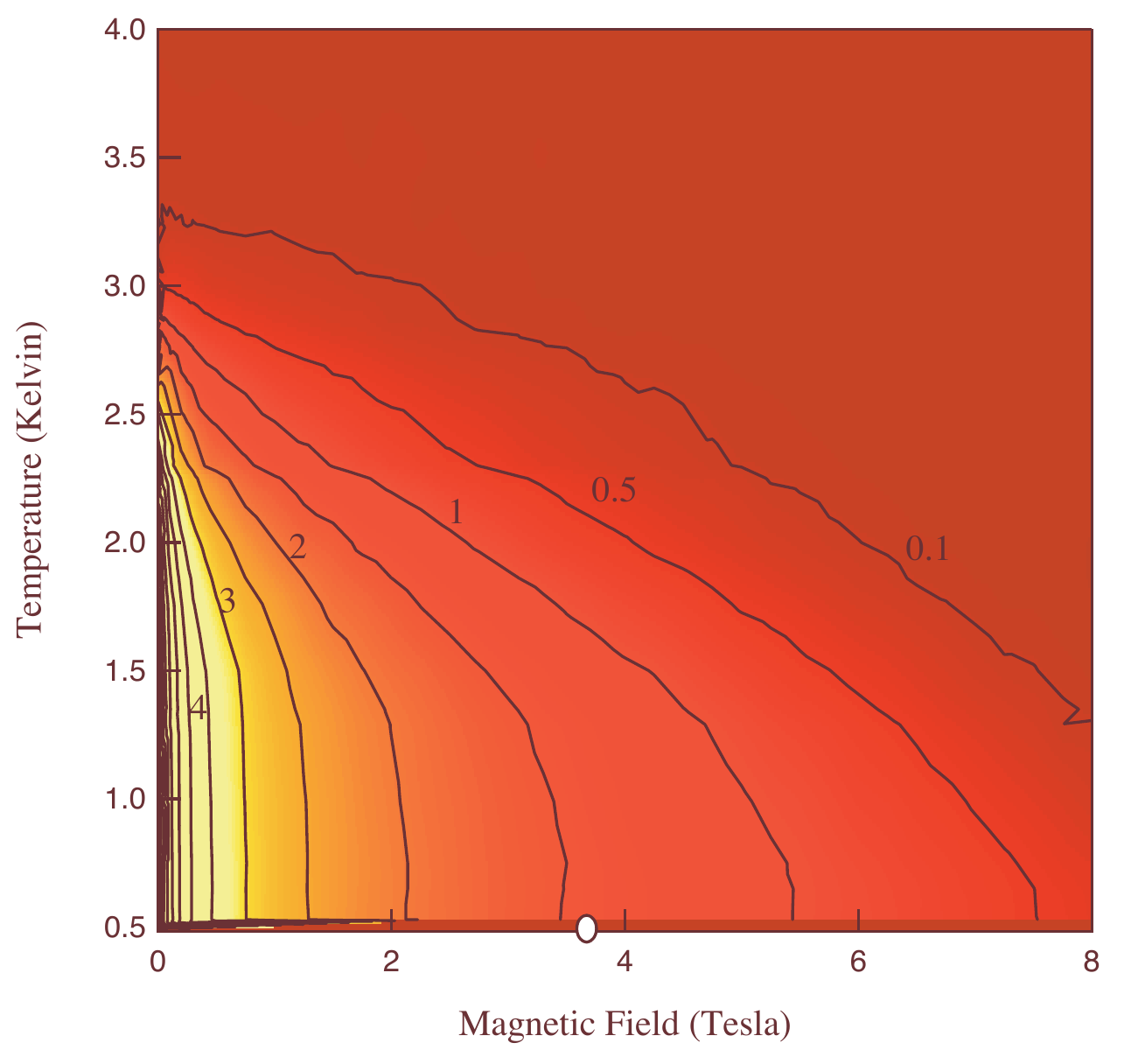}
\end{center}
\caption{ Superfluid stiffness ($\propto \omega \sigma_2$) at 22 GHz given in units of degrees Kelvin.  The critical field, $H_{SIT}$, defined as the iso-resistance point in DC measurements, is shown as a white dot at $H=3.68 T$. Yellow indicates maximum.} \label{fig1}
\end{figure} 

Our observation of a finite frequency superfluid stiffness at $H>H_{SIT}$ is not inconsistent with an insulating T=0 groundstate. As alluded to above, our experiments are sensitive to superfluid \textit{fluctuations} because we probe the system on short time scales via an experimental frequency $\omega_{Exp}$ that is presumably high compared to an intrinsic order parameter fluctuation rate $\omega_{QC}$ close to the transition.  We note that at low temperatures and well into the insulating side of the phase diagram, the superfluid stiffness becomes temperature independent as $T \rightarrow 0$.  This shows that the observed effects are not thermally driven and is indicative of their intrinsic quantum mechanical nature.  Although we cannot rule out inhomogeneous superconducting patches \cite{Nandini}, we consider the large  ($10^6 \Omega/\Box$) and strongly diverging resistance of our samples at the highest fields and low temperatures to at least favor an interpretation of a material which is globally insulating.

\begin{figure}[!ht]
\begin{center}
\includegraphics[angle=0,width=0.4\textwidth]{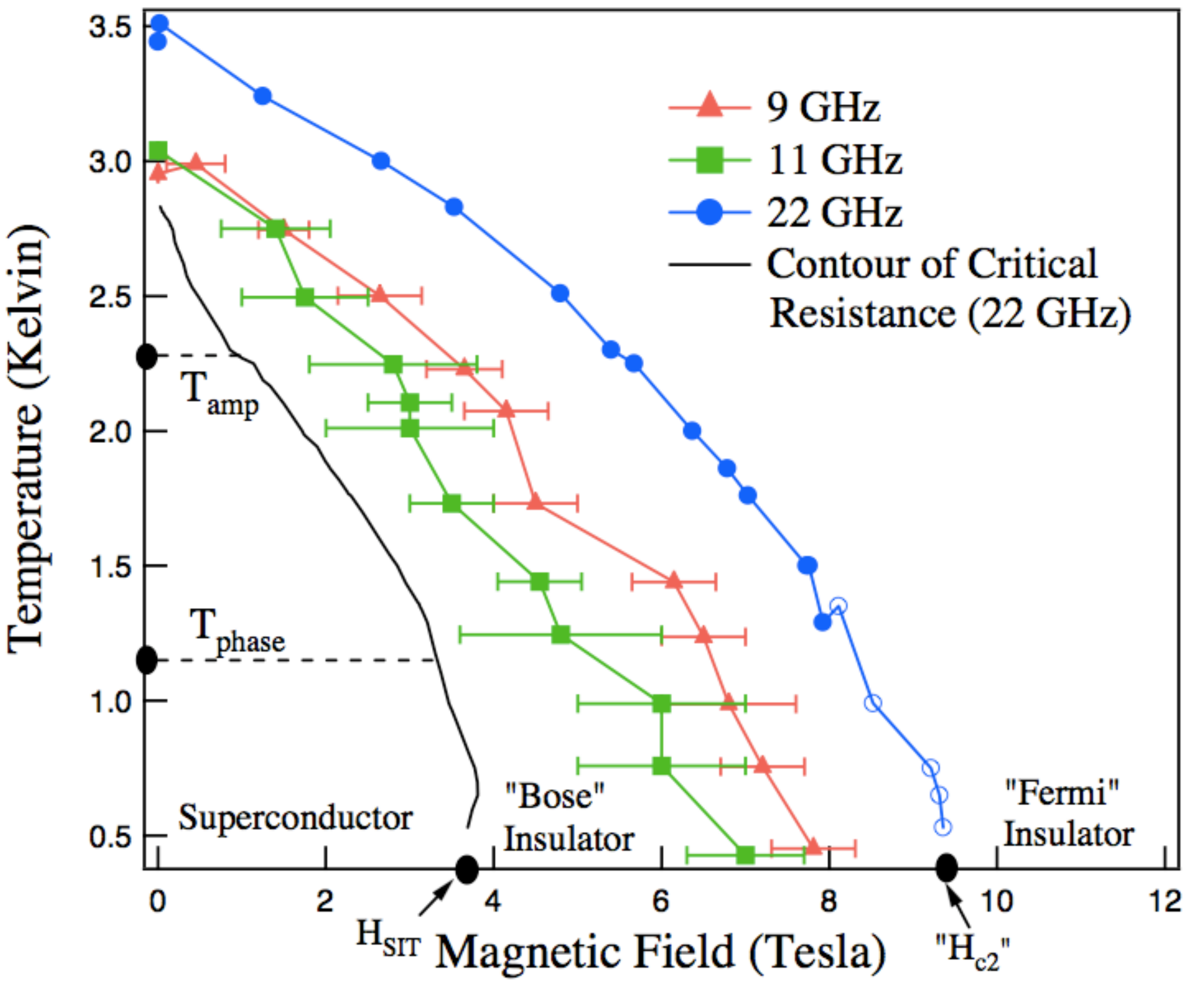}
\end{center}
\caption{2D Field Tuned Superconductor-Insulator ``Phase Diagram''.  Contours with markers are the detection limit for $T_{\theta}$ at specified frequencies. Black dots on the temperature axis denote $T_{amp}$ ($=T_{c0}$) and $T_{phase}$ ($=T_{KTB}$) which are temperatures signifying the onset of amplitude and phase fluctuations, respectively.  They and the contour of critical resistance appearing as a solid black line are described in Ref. \cite{CraneFiniteField}.  $H_{SIT}$ appears as a black dot on the horizontal axis.  Open symbols represent data obtained from a small linear extrapolation beyond our maximum field of 8 Tesla.} \label{fig2}
\end{figure}

We extract a phase diagram that establishes the existence of superconducting correlations well into the insulating state. In Fig. 2 contours are plotted which denote the region above which our superfluid stiffness becomes almost indistinguishable from the normal state noise level (set at 1 \% of the $T \rightarrow 0$, $H=0$ superfluid stiffness), thereby giving a measure of the extent of superconducting correlations into the insulating regime.  Although the noise is greatest for frequency contours away from our cavity's optimal operating frequency of 22 GHz, it is evident that the higher frequencies allows one to examine the fluctuations of the order parameter persisting at fields higher than $H_{SIT}$ at shorter length and time scales than the 9 or 11 GHz probes. In the high-frequency limit one would expect that the detection limit would eventually extrapolate to a field where the Cooper pairs are completely depaired.  Our phase diagram is characterized by a superconducting groundstate, a transition at H$_{SIT}$ to a `Bose' insulating state  with quantum superconducting fluctuations and a crossover to a `Fermi' insulator near a depairing field ``H$_{c2}$''.  We observe superconducting correlations not just asymptotically close to the SIT, but in an extended region above $H_{SIT}$.

\end{document}